\documentclass[pra,reprint,showpacs,superscriptaddress]{revtex4-1}

\usepackage[USenglish]{babel} 

\usepackage{natbib}

\usepackage{booktabs} 

\usepackage{enumitem} 

\usepackage{hyperref} 
\hypersetup{
		colorlinks=true,
		linkcolor=blue,
		citecolor=red,
		urlcolor=blue
}

\usepackage{graphicx}
\usepackage{multirow}
\usepackage{xcolor}
\usepackage{braket}

\usepackage{float}

\usepackage{siunitx}
\usepackage{amsmath}

\begin{document}
\title{Rydberg blockade in an ultracold strontium gas revealed by two-photon excitation dynamics} 

\author{Chang Qiao}
\altaffiliation{These authors contributed equally to this work.}

\affiliation{Hefei National Laboratory for Physical Sciences at the Microscale and Department of Modern Physics, University of Science and Technology of China, Hefei, Anhui 230026, China}
\affiliation{CAS Center for Excellence and Synergetic Innovation Center in Quantum Information and Quantum Physics, University of Science and Technology of China, Shanghai 201315, China}

\author{Can-Zhu Tan}
\altaffiliation{These authors contributed equally to this work.}

\affiliation{Hefei National Laboratory for Physical Sciences at the Microscale and Department of Modern Physics, University of Science and Technology of China, Hefei, Anhui 230026, China}
\affiliation{CAS Center for Excellence and Synergetic Innovation Center in Quantum Information and Quantum Physics, University of Science and Technology of China, Shanghai 201315, China}

\author{Julia Siegl}
\affiliation{Hefei National Laboratory for Physical Sciences at the Microscale and Department of Modern Physics, University of Science and Technology of China, Hefei, Anhui 230026, China}
\affiliation{Physikalisches Institut, Universit\"at Heidelberg, Im Neuenheimer Feld 226, 69120 Heidelberg, Germany}
\author{Fa-Chao Hu}
\author{Zhi-Jing Niu}
\affiliation{Hefei National Laboratory for Physical Sciences at the Microscale and Department of Modern Physics, University of Science and Technology of China, Hefei, Anhui 230026, China}
\affiliation{CAS Center for Excellence and Synergetic Innovation Center in Quantum Information and Quantum Physics, University of Science and Technology of China, Shanghai 201315, China}

\author{Y.H. Jiang}
\email{jiangyh@sari.ac.cn}
\affiliation{Shanghai Advanced Research Institute, Chinese Academy of Sciences, Shanghai 201210, China}
\affiliation{CAS Center for Excellence and Synergetic Innovation Center in Quantum Information and Quantum Physics, University of Science and Technology of China, Shanghai 201315, China}

\author{Matthias Weidem\"uller}
\email{weidemueller@uni-heidelberg.de}
\affiliation{Hefei National Laboratory for Physical Sciences at the Microscale and Department of Modern Physics, University of Science and Technology of China, Hefei, Anhui 230026, China}
\affiliation{CAS Center for Excellence and Synergetic Innovation Center in Quantum Information and Quantum Physics, University of Science and Technology of China, Shanghai 201315, China}
\affiliation{Physikalisches Institut, Universit\"at Heidelberg, Im Neuenheimer Feld 226, 69120 Heidelberg, Germany}

\author{Bing Zhu}
\email{bzhu@physi.uni-heidelberg.de}
\affiliation{Physikalisches Institut, Universit\"at Heidelberg, Im Neuenheimer Feld 226, 69120 Heidelberg, Germany}
\affiliation{CAS Center for Excellence and Synergetic Innovation Center in Quantum Information and Quantum Physics, University of Science and Technology of China, Shanghai 201315, China}

\date{\today} 

\begin{abstract}
We demonstrate the interaction-induced blockade effect in an ultracold $^{88}$Sr gas via studying the time dynamics of a two-photon excitation to the triplet Rydberg series $5\mathrm{s}n\mathrm{s}\, ^3\textrm{S}_1$ for five different principle quantum numbers $n$ ranging from 19 to 37. By using a multi-pulse excitation sequence to increase the detection sensitivity we could identify Rydberg-excitation-induced atom losses as low as $<1\%$. Based on an optical Bloch equation formalism, treating the Rydberg-Rydberg interaction on a mean-field level, the van der Waals coefficients are extracted from the observed dynamics, which agree fairly well with \emph{ab initio} calculations. 
\end{abstract}

\maketitle


\section{Introduction}

The strong Rydberg-Rydberg interactions arising from large electric dipole moments \cite{Gallagher2008} are central to the research frontiers using ultracold Rydberg atoms, including quantum information processing with Rydberg-mediated quantum gates \cite{Saffman2010, Comparat2010, Saffman2016}, Rydberg nonlinear optics \cite{Pritchard2012a, Firstenberg2016}, and Rydberg simulators of many-body physics \cite{Browaeys2016, Browaeys2020}. Recently, there is an increasing interest in extending the studies of Rydberg physics to alkali-earth-like atoms \cite{Dunning2016}. Due to the presence of an extra valence electron compared to the well-studied alkalies \cite{Gallagher2005}, divalent atoms offer Rydberg series of both singlet and triplet characters with richer interactions in a single element, the possibility of co-trapping Rydberg and ground state atoms \cite{Mukherjee2011, Wilson2019, Madjarov2020}, and the allowance of studying two-electron correlations \cite{Kalinski2003}. Alkali-earth-like Rydberg atoms also find applications in realizing spin-squeezing for precision metrology \cite{Gil2014, Kaubruegger2019}, generating high-fidelity entanglement for quantum information processing \cite{Wilson2019, Madjarov2020}, as well as simulating strongly correlated many-body systems \cite{Mukherjee2011}.

Most applications involving Rydberg atoms rely on the \emph{ab initio} calculated Rydberg-Rydberg interaction potentials using for example open-source codes \cite{Weber2017, Robertson2021}, which have been confirmed for alkali atoms by experiments (see like \cite{Gallagher2008, Marcassa2014, Beguin2013}). Specifically, Ref. \cite{Beguin2013} directly extracted the van der Waals (vdW) interaction between two isolated Rydberg atoms by measuring the excitation dynamics and comparing it to solutions of an optical Bloch equation (OBE). The observed suppression of Rydberg excitations at strong interactions is called blockade effect \cite{Lukin2001, Urban2009, Gaetan2009}, which was also studied in larger cold atom ensembles \cite{Singer2004, Tong2004, Heidemann2007, Pritchard2010, Dudin2012, Schauss2012}. Mean-field theories have been applied to account for related observations in bulk atomic gases \cite{Tong2004, Weimer2008, Helmrich2018, DeHond2018}. 

The \emph{ab initio} calculations of Rydberg properties in alkali-earth-like systems are much more complicated than that for alkali atoms due to the existing of two-electron effects \cite{Vaillant2012, Ye2013a, Vaillant2014, Robicheaux2019}. Although the direct method of measuring the vdW interactions like Ref. \cite{Beguin2013} provides the most reliable benchmark for these calculations, the experiment requires a challenging setup for being able to control the distance between two isolated single atoms in a large scale ($1\sim20$~$\mu$m). So far, the experimental studies of the Rydberg-Rydberg interactions in alkali-earth-like atoms are quite limited, where the blockade effect was studied spectroscopically in strontium \cite{Zhang2015, DeSalvo2016} and high-fidelity entanglement was generated via the strong Rydberg-Rydberg interactions in an optical tweezer experiment \cite{Madjarov2020}. 

\begin{figure*}[t!]
	\centering
	\includegraphics[angle=0,width=0.9\textwidth]{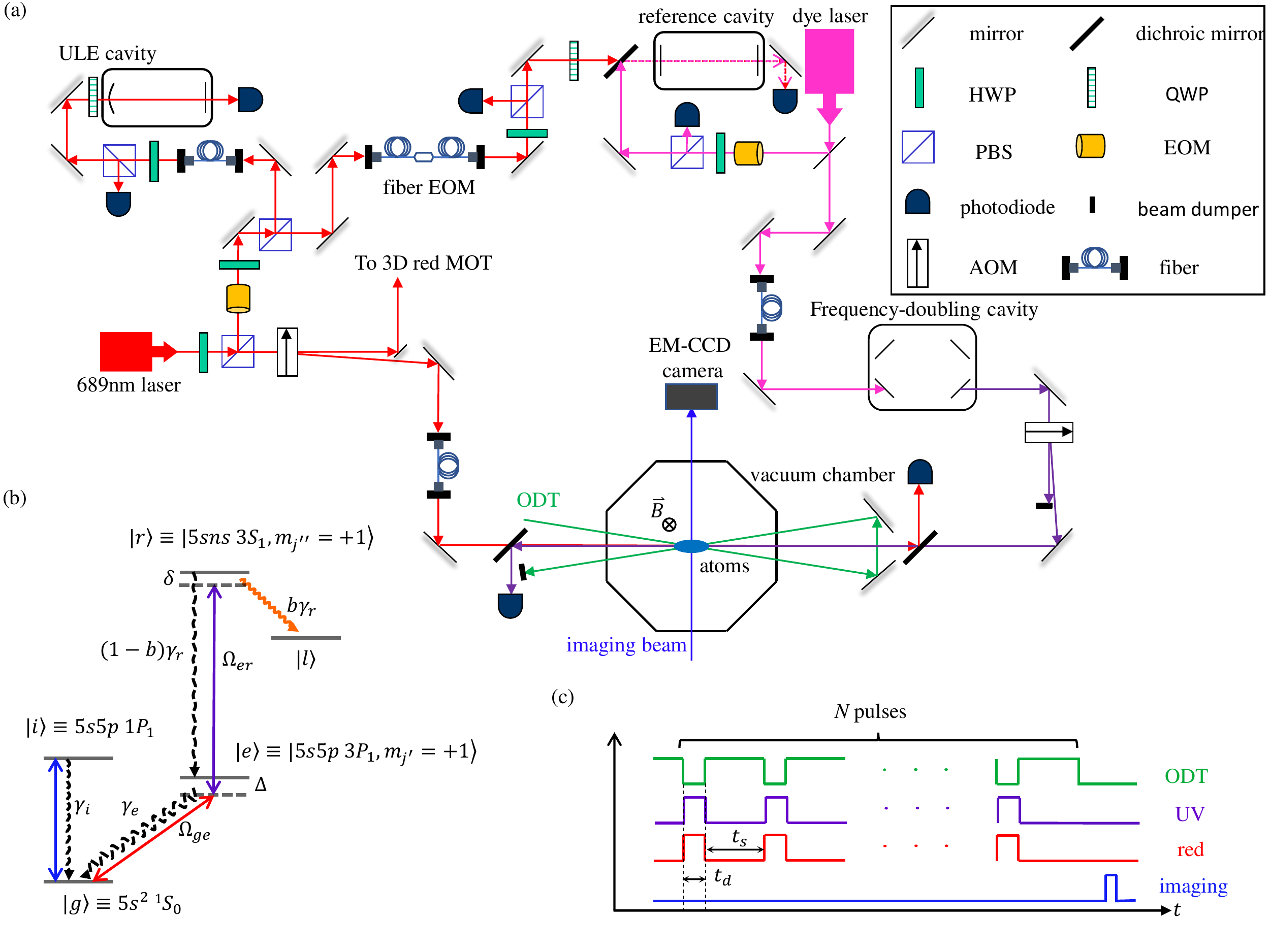}
	\caption{Two-photon pulsed excitation to the $5sns\text{ }^3S_1$ states. (a) Schematic view of the experimental setup. The abbreviations are: HWP\textemdash half wave-plate, QWP\textemdash quarter wave-plate, PBS\textemdash polarizing beam-splitter, EOM\textemdash electro-optic modulator, AOM\textemdash accousto-optic modulator, ULE\textemdash ultra-low expansion. Many optical elements are omitted here, like lenses. See text for detailed descriptions. (b) The relevant energy levels. The Rydberg state $\ket{r}$ is addressed via a two-photon process from the ground state $\ket{g}$ via an intermediate state $\ket{e}$. $\Omega_{er}$ ($\Omega_{ge}$) represents the Rabi frequency driving the $\ket{e}-\ket{r}$ ($\ket{g}-\ket{e}$) transition and $\Delta$ ($\delta$) is the one-photon (two-photon) detuning. Atoms at $\ket{g}$ are detected by absorption imaging with the broad $\ket{g}-\ket{i}$ transition. $\gamma_{g, e, i}$ are the natural linewidths of the corresponding states and the branching ratio from $\ket{r}$ to $\ket{e}$ ($\ket{l}$) is $1-b$ ($b$). Here $\ket{l}$ represents the low-lying states that $\ket{r}$ decays to, except $\ket{e}$. (c) The $N$-pulse sequence. The two beams are operated simultaneously with $N$ pulses, each of equal duration $t_{d}$ and equally separated by $t_s$. The ODT is temporarily switched off during the excitation pules.}
	\label{fig:exp}
\end{figure*}

In this work we present a systematic study of the Rydberg blockade effect in an ultracold gas of $^{88}$Sr atoms using two-photon excitations from the ground state $5s^2\text{ }^1S_0$ ($\equiv\ket{g}$) to the triplet Rydberg series $5sns\text{ }^3S_1$ ($\equiv\ket{r}$) via the long-lived $5s5p\text{ }^3P_1$ state ($\equiv\ket{e}$), where five different principle quantum numbers $n$ in the range of $19\le n\le37$ are addressed. By making use of a multi-pulse scheme to increase the detection sensitivity \cite{DeSalvo2016}, we could identify Rydberg-excitation-induced atom losses smaller than $1\%$. The suppression of the Rydberg excitations is observed by comparing the dynamics at two different atomic densities. The measurements are modelled by a single-particle OBE taking the Rydberg-Rydberg interaction into account with a self-consistent mean-field theory \cite{Helmrich2018, DeHond2018}, from which the vdW interaction coefficients $C_6$ are extracted and compared to the \emph{ab initio} calculations in Ref. \cite{Vaillant2012}. Despite the simplicity of our model, a fairly good agreement is found between the measured and calculated $C_6$ coefficients.

The article is organized as follows: The experimental details are presented in Sec. \ref{sec: exp} and the mean-field model is introduced in Sec. \ref{sec:model}. The experimental results and the mean-field modelling are discussed in Secs. \ref{sec:noninteracting} and \ref{sec:interacting}, from which the vdW coefficients are extracted and compared to theoretical calculations in Sec. \ref{sec:vdW}. We conclude the paper with Sec. \ref{sec:conclusion}. 

\section{Methods}
\subsection{Experimental Setup}
\label{sec: exp}

Our experimental setup for producing ultracold strontium gases via a two-stage magneto-optical cooling and trapping has been described previously in Refs. \cite{Nosske2017, Hu2019, Qiao2019a}. In brief, a three-dimension (3D) magneto-optical trap (MOT) operating on the broad $\ket{g} \rightarrow 5s5p\text{ }^1P_1$ ($\equiv\ket{i}$) transition at \SI{461}{nm} is loaded from a two-dimension (2D) MOT using the same transition \cite{Nosske2017}. Atoms in this broadband MOT can decay to the $5s5p\text{ }^3P_2$ state, which is long-lived and magnetically trappable \cite{Nagel2003}. Then, via a single-frequency repumping at \SI{481}{nm} \cite{Hu2019}, atoms accumulated in the MOT magnetic quadrupole field for \SI{5}{s} are transferred to a second 3D MOT (red MOT) operating on the narrow $\ket{g} \rightarrow \ket{e}$ transition with a natural linewidth of $\gamma_e = 2\pi\times$ \SI{7.5}{kHz}. The red MOT stage lasts for about \SI{200}{ms} \cite{Qiao2019a} and a crossed optical dipole trap (ODT) at \SI{532}{nm} is turned on \SI{150}{ms} before switching off the red MOT. Then the atoms are held in the ODT for certain time before applying Rydberg excitation pulses. The two ODTs cross at an angle of \SI{18}{\degree} [see Fig. \hyperref[fig:exp]{1(a)}], which results in an elongated atomic cloud with an aspect ratio of about $3:1$ ($\sim$\SI{25}{\mu m} in the radial direction). Due to the very small collisional rates in the gas of  ground state $^{88}$Sr atoms \cite{Yan2011}, the atomic density monotonically decays during the holding in the ODT while the temperature and cloud size remain almost unchanged. The resulting atomic densities are listed in Table \ref{tab:interaction} at holding times of \SI{100}{ms} and \SI{1000}{ms} for various principle quantum numbers.

The Rydberg states $\ket{r}$ are addressed via a two-photon process [see Fig. \hyperref[fig:exp]{1(b)} for the corresponding level structures]. The optical setup for generating and controlling the two excitation beams is schematically shown in Fig. \hyperref[fig:exp]{1(a)}. The 689-nm red beam driving the $\ket{g}-\ket{e}$ transition is delivered from an external cavity diode laser, which is frequency locked to a passive ultra-stable cavity resulting in a linewidth smaller than \SI{10}{kHz} \cite{Qiao2019a}. The upper transition ($\ket{e}-\ket{r}$) is driven by a UV beam at $\sim$\SI{319}{nm}, which is obtained by frequency doubling a dye laser using resonant cavity second harmonic generation \cite{Arias2017}. The commercial dye laser (Matisse 2 DX, Sirah) is locked to a low-finesse ($\sim700$) reference cavity to reduce the linewidth (below \SI{20}{kHz} over \SI{1}{ms} \cite{Couturier2018}). Via a standard PDH setup, the length of this low-finesse cavity is stabilized to the frequency-locked 689-nm laser, between which a fiber electro-optic modulator (EOM) is inserted. Since one of the first-order modulation sidebands is used in the stabilization, output frequency of the dye laser can be tuned by the EOM driving frequency to cover the free spectral range of the reference cavity ($\sim$\SI{1.2}{GHz}). The power of both excitation beams are actively stabilized before the last AOMs for switching and monitored after passing through the atomic cloud with two photodiodes [see Fig. \hyperref[fig:exp]{1(a)}]. 

The two excitation beams counter-propagate horizontally along the axial direction of the atomic cloud [see Fig. \hyperref[fig:exp]{1(a)}]. The red (UV) beam is linearly polarized in the horizontal (vertical) direction with a $1/e^2$-diameter of \SI{1.4}{mm} (\SI{1.2}{mm}). A quantization field of about \SI{1.16}{G} is applied vertically, giving rise to a Zeeman splitting of about $h\times$\SI{4.86}{MHz} (\SI{8.10}{MHz}) for the $\ket{e}$ ($\ket{r}$) state. The frequency of the 689-nm light is tuned close to the $\sigma^{+}$ transition ($\Delta m_j=+1$) such that $\ket{e}$ refers to the $m_{j'}=+1$ level of the $5s5p\text{ }^3P_1$ state hereafter. Here $m_{j}$ are the magnetic quantum numbers. The vertically polarized UV beam drives only the $\pi$ transition ($\Delta m_j=0$), $\ket{r}$ also refers to the $m_{j''}=+1$ level of Rydberg states. As seen in Fig. \hyperref[fig:exp]{1(b)}, the one-photon (two-photon) detuning is defined as $\Delta=\omega_\textrm{red}-(\omega_e-\omega_g)$ [$\delta=\omega_\textrm{UV}+\omega_\textrm{red}-(\omega_r-\omega_g)$], where $\omega_\textrm{red}$ ($\omega_\textrm{UV}$) is the angular frequency of the red (UV) excitation light and $\hbar\omega_{g, e, r}$ are the energies of corresponding levels. A loss state $\ket{l}$ is introduced to represent all the lower-lying states, that the Rydberg state $\ket{r}$ decays to, except $\ket{e}$. The branching ratio $b$ from $\ket{r}$ into $\ket{l}$ can be calculated to be $2/3$ using the Wigner-Eckart theorem \cite{DeSalvo2016}.

The excitation sequence is schematically illustrated in Fig. \hyperref[fig:exp]{1(c)}. The two excitation beams are pulsed with $N$ equally spaced (of $t_s$) pulses of a duration $t_d$. To avoid the possible differential AC Stark effects on both red and UV transitions, the ODT is also pulsed simultaneously with the excitation beams. We keep $t_s$ long enough when compared to the lifetimes of $\ket{r}$ and $\ket{e}$, such that all the atoms populate $\ket{g}$ at the beginning of each pulse. The remaining ground state atom number after the pulse sequence is recorded by absorption imaging on the broad $\ket{g} \rightarrow \ket{i}$ transition.

\subsection{The mean-field model}
\label{sec:model}

Describing the Rydberg-Rydberg interaction on a mean-field level, the evolution of the system within a single excitation pulse can be modelled by a single-particle OBE with dephasing effects in the Lindblad form ($\hbar=1$) \cite{DeSalvo2016, Helmrich2018, DeHond2018}
\begin{equation}
\label{eq:obe}
\dot{\rho}=-i[H, \rho]+\sum\limits_{i}L_i\rho L_i^\dagger - \frac{1}{2}\{ L_i^\dagger L_i, \rho\} \, .
\end{equation}
Here $\rho$ is the density matrix for a four-level system of $\ket{g}$, $\ket{e}$, $\ket{r}$, and $\ket{l}$ (see Fig. \hyperref[fig:exp]{1(b)}), and $[\cdots]$ ($\{\cdots\}$) denotes the (anti-)commutator. We refer to the matrix element of $\rho$ as $\rho_{ab}\ket{a}\bra{b}$ with $a, b \in \{g, e, r, l\}$. In the experiment, we monitor the remaining atom number fraction in the ground state $\ket{g}$ after the excitation, which can be represented as
\begin{equation}
\label{eq:losefraction}
1-p_d=1-\rho_{ll}-b\rho_{rr}.
\end{equation} 
Here $p_d$ is the loss fraction of a pulse of length $t_d$, including all the atoms populated the loss state $\ket{l}$ during and after the single-pulse excitation. 

\begin{figure*}[t]
	\centering
	\includegraphics[width=\textwidth]{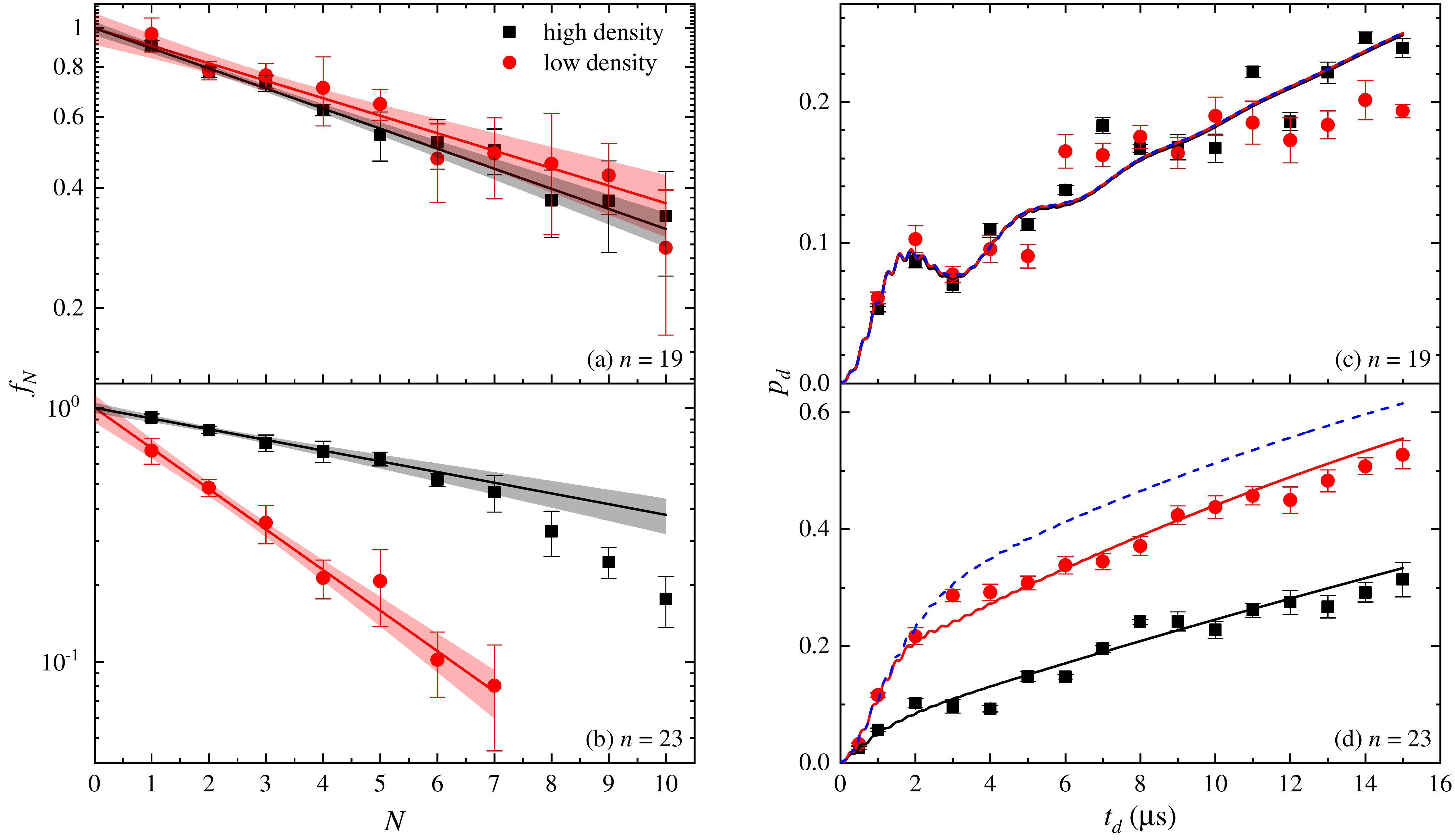}
	\caption{Measurements of the loss fraction per pulse $p_d$ for $n=19$ and $n=23$. (a, b) The remaining fraction of atoms $f_N$ as a function of the pulse number $N$ for these two states at two different atomic densities (see Table \ref{tab:interaction}). The solid curves are fits to Eq. \eqref{eq:remainfrac} to extract $p_d$ and the shaded areas mark the $95\%$ confidence level. See text for detailed discussions. (c, d) $p_d$ as a function of the single-pulse length $t_d$. In (c), the black, red, and blue curves are parameter-free calculations for the $n=19$ state with $V/h=0.62$, 0.10, and 0 (see text), respectively. In (d), the solid curves are fits of the high- and low-density data to the mean-field model, from which we obtain $\gamma_{er}/2\pi=$~\SI{0.98(36)}{MHz}, $V^e/2\pi=$~\SI{-71(10)}{MHz} for the high-density measurement (black curve), and $\gamma_{er}/2\pi=$~\SI{0.10(9)}{MHz}, $V^e/2\pi=$~\SI{-3.4(7)}{MHz} for the low-density data (red curve). The fitted parameters are listed in Table \ref{tab:interaction}. As a comparison, the blue dashed line represents the non-interacting case ($V=0$) with $\gamma_{er}/2\pi=$~\SI{0.10}{MHz}.}
	\label{fig:23pn}
\end{figure*}

The Hamiltonian $H$ governing the two-photon excitation dynamics \cite{Helmrich2018, DeHond2018} reads
\begin{equation}\label{eq:hamiltonian}
\begin{aligned}
H = & (\frac{\Omega_{ge}}{2}\ket{g}\bra{e} + \frac{\Omega_{er}}{2}\ket{e}\bra{r}+h.c.) \\ & - (\delta - V\rho_{rr}^2)\ket{r}\bra{r} - \Delta\ket{e}\bra{e} \, ,  
\end{aligned}
\end{equation}
where the Rabi frequencies and detunings are as shown in Fig. \hyperref[fig:exp]{1(b)} and $\rho_{rr}$ is the Rydberg fraction. We have omitted the diagonal term with $\ket{l}$ since it plays no role in determining the population dynamics. The characteristic interaction energy $V$ reads
\begin{equation}\label{eq: Interaction}
	\begin{aligned}
		V = C_6(\frac{4\pi}{3}n_0)^2 \, .
	\end{aligned}
\end{equation}
Here $C_6$ is the coefficient for the vdW interaction between atoms in $\ket{r}$ and $n_0$ is the ground state atomic density. This mean-field form is derived self-consistently in Ref. \cite{Helmrich2018}. 

The spontaneous decays of $\ket{e}$ and $\ket{r}$ are described by quantum jump operators $L_e=\sqrt{\gamma_e}\ket{g}\bra{e}$ and $L_r=\sqrt{(1-b)\gamma_r}\ket{e}\bra{r}+\sqrt{b\gamma_r}\ket{l}\bra{r}$,
where $b=2/3$ is the branching ratio from $\ket{r}$ to $\ket{l}$ \cite{Footnote1}. Effects like laser frequency noises, magnetic field fluctuations, and Doppler effects, cause the fluctuations of the detunings and lead to dephasing, especially the two-photon detuning $\delta$, can be accounted for by additional jump operators in the form of $L_{er}=\sqrt{\gamma_{er}}\ket{r}\bra{r}$ and $L_{ge}=\sqrt{\gamma_{ge}}\ket{e}\bra{e}$. $\gamma_{ge}$ is characterized by measuring the damped Rabi oscillations between $\ket{g}$ and $\ket{e}$ (see Appendix \ref{sec:calib}). 

\section{Results}
\label{sec:res}

\subsection{The noninteracting case}
\label{sec:noninteracting}

\begin{table*}[t!]
	\caption{Rydberg-Rydberg interaction. $n_0$ are the atomic densities with determination uncertainties in the corresponding brackets. Two free parameters, the mean-field interaction energy $V^{e} = C_6^{e}(\frac{4\pi}{3}n_0)^2$ and the dephasing rate $\gamma_{er}$, are extracted from experiments and listed (see text for details). The numbers in brackets are the fitting errors from the measurement statistics, while those in super- and sub-scripts are systematic uncertainties resulting from the Rydberg lifetime ($1/\gamma_r$) uncertainty \cite{Kunze1993}.}
	\label{tab:interaction}
	\begin{ruledtabular}
		\begin{tabular}{lcccc}
			\multirow{2}{2em}{$n$} & $n_0$ & $V^e/h$ & $\gamma_{er}/2\pi$ & Rydberg lifetime ($1/\gamma_r$) \cite{Kunze1993} \\
			& $10^{12}$~cm$^{-3}$ & MHz & MHz & $\mu$s \\
			\hline
			\multirow{2}{2em}{$19$} & 1.47(21)  & \multirow{2}{2em}{$0^{+4.2}$} & \multirow{2}{2em}{0.03(1)} &  \multirow{2}{5em}{$1.83\pm0.20$} \\
			& 0.58(13) &  & & \\ 
			\hline
			\multirow{2}{2em}{$23$} & 1.14(19) & $71(10)^{+31}_{-18}$ & $0.98(36)^{+0.06}_{-0.08}$ & \multirow{2}{5em}{$3.36\pm0.71$} \\
			& 0.40(7) & $3.4(7)^{+1.9}_{-1.1}$ & $0.10(9)^{+0.01}_{-0.03}$ & \\
			\hline
			\multirow{2}{2em}{$27$} & 1.52(25) & $226(18)^{+63}_{-37}$ & $0.70(9)^{+0.06}_{-0.27}$ & \multirow{2}{5em}{$5.9\pm1.2$}\\
			& 0.62(9) & $48(4)^{+21}_{-12}$ & $0.31(9)^{+0.02}_{-0.03}$ & \\
			\hline
			\multirow{2}{2em}{$30$} & 0.88(25) & $689(94)^{+174}_{-101}$ & $1.13(31)^{+0.02}_{-0.05}$ & \multirow{2}{5em}{$8.4\pm1.8$}\\
			& 0.33(10) & $31(14)^{+17}_{-13}$ & $1.31(12)^{+0.04}_{-0.07}$ & \\
			\hline
			\multirow{2}{2em}{$37$} & $0.95(11)$ & $2870(194)^{+446}_{-240}$  & $2.41(37)^{+0.19}_{-0.30}$ & \multirow{2}{5em}{$16.9\pm3.6$} \\
			& 0.36(4) & $321(31)^{+61}_{-38}$ & $1.47(16)^{+0.07}_{-0.10}$ & 
		\end{tabular}
	\end{ruledtabular}
\end{table*}

Figure \hyperref[fig:23pn]{2(a)} shows the measurement of the fractional ground state atom number $f_N$ as a function of the applied pulse number $N$ with $t_d=$~\SI{4}{\mu s} and $t_s=$~\SI{50}{\mu s} [see Sec. \ref{sec: exp} and Fig. \hyperref[fig:exp]{1(c)}] at two different atomic densities for $n=19$ (see Table \ref{tab:interaction}). The Rabi frequencies $\Omega_{ge}$ and $\Omega_{er}$ are calibrated to be $2\pi\times$~\SI{813(20)}{kHz} and $2\pi\times$~\SI{862(40)}{kHz} by driving Rabi oscillations between $\ket{g}$ and $\ket{e}$ state and measuring the Aulter-Townes splitting (ATS) of $\ket{e}$ caused by the UV light, respectively. The details of these calibrations are presented in the Appendix \ref{sec:calib}. The one-photon (two-photon) detuning $\Delta$ ($\delta$) is $2\pi\times$\SI{3}{MHz} ($2\pi\times$\SI{0.27}{MHz}). 

Since $t_s$ is long compared to the lifetimes of both $\ket{r}$ $\{1/\gamma_r=\SI{1.83(20)}{\mu s}$ \cite{Kunze1993}$\}$ and $\ket{e}$ ($1/\gamma_e\sim$\SI{21}{\mu s}), we can safely assume that all the atoms are at the ground state at the beginning of each pulse. The fraction of lost atoms $p_d$ due to a single pulse is a constant if the interaction $V$ is unchanged or zero during the measurement and the remaining atom fraction $f_N$ as a function of $N$ is expressed as 
\begin{equation}
	\label{eq:remainfrac}
	f_N=(1-p_d)^N ,
\end{equation}
which is used to fit the experiment data in Figs. \hyperref[fig:23pn]{2(a)} and \hyperref[fig:23pn]{(b)} to extract the corresponding loss fraction per excitation pulse $p_d$. In Fig. \hyperref[fig:23pn]{2(c)} we show the measurement of $p_d$ with different pulse lengths $t_d$. Similar $p_d$ are observed for all the data points at two densities differed by a factor of about 2.5 (see Table \ref{tab:interaction}). Eq. \eqref{eq:obe} is numerically solved with and without including the vdW interactions to obtain $p_d$ from Eq. \eqref{eq:losefraction}, which are also shown in Fig. \hyperref[fig:23pn]{2(c)}. Here in the numerical calculations the dephasing term $\gamma_{ge}$ is determined experimentally (see Appendix \ref{sec:calib}), $\gamma_{er}$ is set to zero, and the $C_6$ coefficient is from Ref. \cite{Vaillant2012}. By comparing the experimental measurements and the numerical results we conclude that the vdW interactions are negligible for the observed excitation dynamics at $n=19$. However, an upper bound of the interaction $V$ can still be obtained, as will be discussed later in Sec. \ref{sec:interacting} and also shown in Table \ref{tab:interaction}. 

\subsection{The interacting case}
\label{sec:interacting}

An example measurement of $f_N$ for the $n=23$ state with $t_d=$ \SI{4}{\mu s} and $t_s=$ \SI{50}{\mu s}, similar as that for $n=19$ in Fig. \hyperref[fig:23pn]{2(a)}, is shown in Fig. \hyperref[fig:23pn]{2(b)}. The Rabi frequencies $\Omega_{ge}$ and $\Omega_{er}$ are $2\pi\times$~\SI{784(5)}{kHz} and $2\pi\times$~\SI{1225(7)}{kHz}, respectively, and $\Delta$ ($\delta$) is $2\pi\times$\SI{3}{MHz} ($-2\pi\times$\SI{0.09}{MHz}). We fit the experimental data in \hyperref[fig:23pn]{2(b)} to Eq. \eqref{eq:remainfrac}. At the low density all the measured points fit very well to the equation, while at the high density only data up to $N=5$ is used in the fitting due to the non-negligible change of $V$ after significant decreasing of atomic densities due to atom losses. The measurements at larger $N$ showing heavier losses agree with the fact that Rydberg excitations are less suppressed at smaller atomic densities.

Similar measurements of $p_d$ for $n=23$ as those in Fig. \hyperref[fig:23pn]{2(c)} are also performed and presented in Fig. \hyperref[fig:23pn]{2(d)}. We observe a much smaller $p_d$ at the higher atomic density, especially with longer pulses. To extract quantitative information about the vdW interactions, we fit the data in Figs. \hyperref[fig:23pn]{2(c)} and \hyperref[fig:23pn]{2(d)} to numerical solutions of Eq. \eqref{eq:obe}, the implementation of which is benchmarked with previous measurements in Refs. \cite{DeSalvo2016, DeHond2018} (see Appendix \ref{sec:beachmark}). The fit is done by minimizing the following cost function
\begin{equation}
\label{eq:costf}
\chi=\sum\limits_i [p_d^t(t_i)-y_i]^2/\sigma_i^2 ,
\end{equation}
where $y$ and $\sigma$ are the measured $p_d$ and their corresponding uncertainties [see Figs. \hyperref[fig:23pn]{2(c)} and \hyperref[fig:23pn]{2(d)}]. The theoretical loss fraction per pulse $p_d^t$ is obtained from Eq. \eqref{eq:losefraction} by numerically solving Eq. \eqref{eq:obe} for certain pulse length $t_d$. For each $n$ and each density, the interaction parameter $V$ and the dephasing rate $\gamma_{er}$ are the two fitting parameters, while others are either experimentally determined or from existing literatures (see Appendix \ref{sec:calib}). As seen in Fig. \hyperref[fig:23pn]{2(d)}, the experimental data can be fitted well by the mean-field model (solid curves) with the obtained fitting parameters ($V^e$ and $\gamma_{er}$) listed in Table \ref{tab:interaction}. The numbers in brackets indicate the fitting errors coming from the error bars shown in Fig. \hyperref[fig:23pn]{2(d)}. 

In the above fitting procedure, the dominating systematic uncertainties (indicated by super- and sub-scripts for the fitting parameters in Table \ref{tab:interaction}) are from the Rydberg-state lifetimes $1/\gamma_r$, which are taken or scaled from Ref. \cite{Kunze1993}. For $n=19$ and $23$ the lifetimes measured there are directly used, while for higher $n$ they are scaled from the lifetime of $n=23$ by assuming $\gamma_r\propto (n-\delta n)^{-3}$ with the quantum defects $\delta n$ from Ref. \cite{Couturier2019}. Here we only take the radiative lifetimes into consideration since the blackbody-radiation-induced transitions to neighbouring levels may still contribute to the blockade effect. We are aware of the lifetime measurement of $n=35$ in Ref. \cite{Kunze1993} and $n=38$ in Ref. \cite{Camargo2016} and the scaled lifetime of $n=37$ in Table \ref{tab:interaction} agrees better with the latter.

In total, the same measurements as those presented in Fig. \ref{fig:23pn} are performed for five different principle quantum numbers of $n=19, 23, 27, 30, 37$. The data is fitted to the mean-field model in Sec. \ref{sec:model} following the procedure outlined above and all the fitted parameters are listed in Table \ref{tab:interaction}. For the $n=19$ state, an upper bound of the interaction energy $V^e$ is given instead of a fitted value. Larger interaction energies $V^e$ are obtained for higher densities and larger $n$. In most cases, the dephasing rate $\gamma_{er}$ is larger when the interaction is stronger, indicating a interaction-induced dephasing \cite{DeSalvo2016}.

\subsection{The vdW coefficients} \label{sec:vdW}

\begin{figure}[t]
	\centering
	\includegraphics[width=0.5\textwidth]{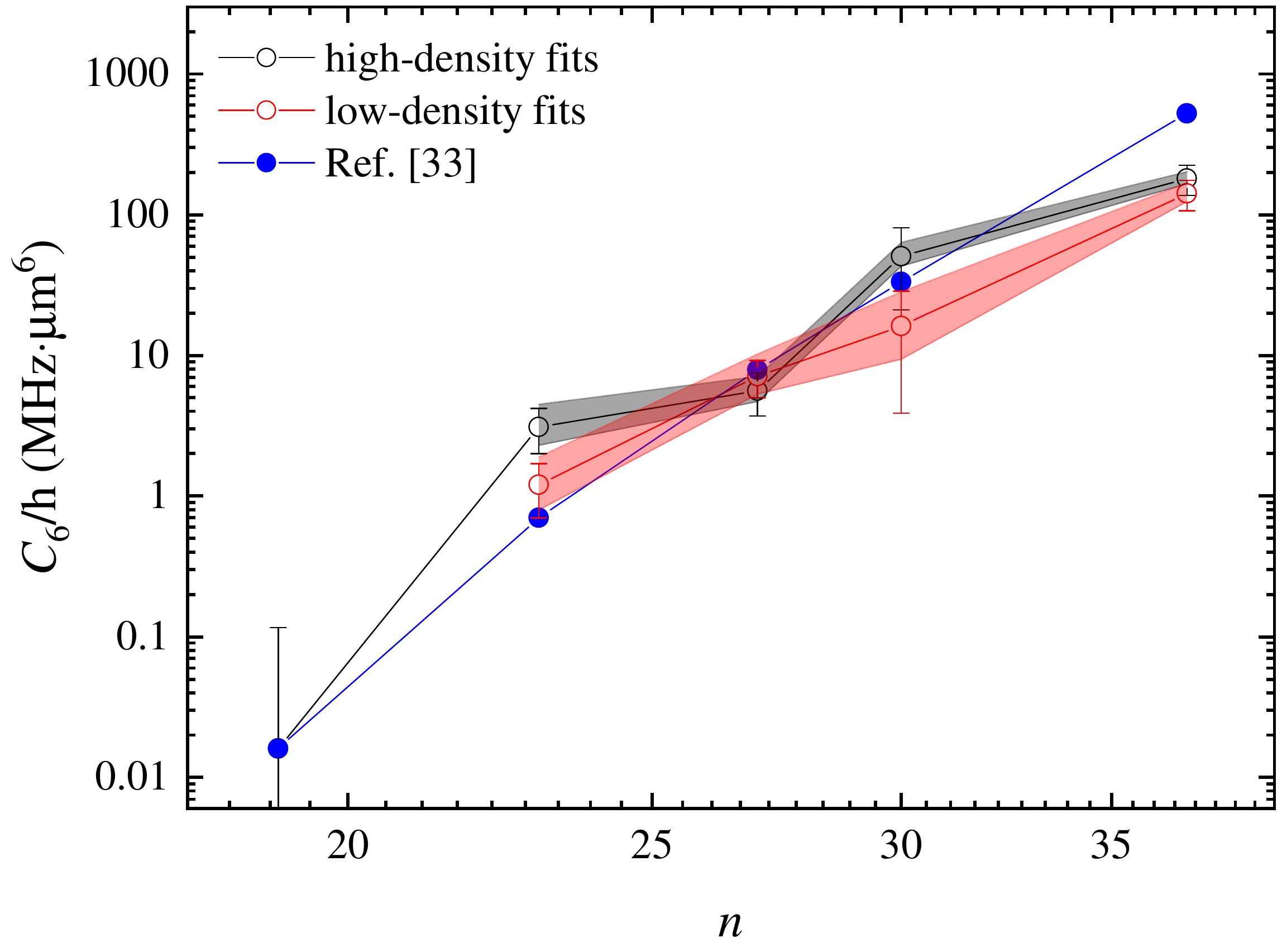}
	\caption{The vdW coefficients $C_6$. The black and red points are from fits to the experimental data at high and low densities, respectively, while the blue ones are from Ref. \cite{Vaillant2012}. See text for more details.}
	\label{fig:C6}
\end{figure}

Via the relation in Eq. \eqref{eq: Interaction}, we extract the vdW coefficients $C_6$ for the measured principle quantum numbers $n$ from the fitted characteristic interaction energy $V^e$ and compare them in Fig. \ref{fig:C6} to the values calculated in Ref. \cite{Vaillant2012} (blue points). For each $n$ (except $n=19$), two $C_6$ values are obtained from the high- (black circles) and low-density (red circles) measurements, respectively. The error bars represent the uncertainties coming from the errors in fitting $V^e$ and determining the atomic densities, both of which are indicated as numbers in brackets in Table \ref{tab:interaction}. The shaded areas mark the uncertainties of $C_6$ resulting from the systematic errors of $V^e$ when performing the fits (see Table \ref{tab:interaction} and Sec. \ref{sec:interacting}).

The dependence of fitted $C_6$ values on $n$ agree qualitatively with the theoretical calculations based on the single-active-electron (SAE) approximation \cite{Vaillant2012}, which is appropriate for high-$n$ states ($n\ge30$ in Ref. \cite{Vaillant2012}). The blue points in Fig. \ref{fig:C6} for the three $n<30$ states are direct extrapolations of the calculation. The largest deviations between experimental and theoretical values happen at $n=23$ and $n=37$. For $n=23$, the fitted $C_6$ from the high-density (low-density) data is a factor of 4 (2) larger than the extrapolated value, which may be due to the inaccuracy of the SAE model for lower-$n$ states. For $n=37$, the measured $C_6$ is smaller than the SAE calculation by a factor of 3. This may be caused by the neglect of effects like level crossings with other molecular potentials and the higher-order terms in the multipole expansion using the current mean-field model, which play more important roles with stronger interactions at higher $n$.

However, we also note that larger uncertainties of the measured $C_6$ could be possible due to imperfections of the experiments as well as possibly larger uncertainties of Rydberg lifetime determination. For example, we have checked laser frequency drifts in our modelling (not included in Fig. \ref{fig:C6}), especially that of the UV laser affecting the two-photon detuning $\delta$. A drift of $2\pi\times$~\SI{100}{kHz} in $\delta$ would lead to a change of about $15\%$ in the fitted $V$ for the $n=23$ high-density measurement (see Table \ref{tab:interaction}). This effect is more significant when the drifting is not small compared to the interaction energy, i.e. the low-density measurements at smaller $n$. We also cannot rule out the effect of any possible residual DC electric fields in our setup, which can lead to smaller $C_6$ values \cite{Ravets2015}. The used Rydberg lifetimes in our model are scaled from Ref. \cite{Kunze1993} and even larger uncertainties could happen. As mentioned in the second last paragraph of Sec. \ref{sec:interacting}, disagreements are observed among the scaled $n=37$ lifetime in Table \ref{tab:interaction}, the $n=35$ measured in Ref. \cite{Kunze1993}, and that for $n=38$ in Ref. \cite{Camargo2016}. Significant improvements could be obtained by having better control on the experimental parameters and performing more precise calculations or measurements on the Rydberg lifetimes.

\section{Conclusion} \label{sec:conclusion}
In conclusion, we have shown the blockade effect in two-photon excitations of strontium triplet Rydberg states by directly measuring the density-dependent excitation dynamics. Using a mean-field model we have extracted the vdW interaction coefficients $C_6$ from the measured dynamics and compared the results to theoretical calculations. Considering the simplicity of our model, the agreement between the extracted and calculated $C_6$ values is fairly good. Such a method could be the first option to investigate the Rydberg physics, when the Rydberg-Rydberg interaction potentials are not well-known for the studied atomic species and no advanced experimental setup like optical tweezers \cite{Beguin2013} is available. However, improvements from both the experimental and theoretical sides are needed to mitigate the still existing experiment-theory deviations.      
 
\section*{Acknowledgements}
We are grateful to Thomas C. Killian, Julius de Hond, and Klaasjan van Druten for sharing their data. We acknowledge L. Couturier, I. Nosske and P. Chen for their contributions at the early stage of this study. We are supported by the Anhui Initiative in Quantum Information Technologies. Y.H.J. also acknowledges support from the National Natural Science Foundation under Grant No. 11827806.

\appendix

\section{Calibrations of experimental parameters}
\label{sec:calib}

\subsection{Rabi oscillations between $\ket{g}$ and $\ket{e}$}

The Rabi frequency $\Omega_{ge}$ is calibrated by directly monitoring the Rabi oscillation between $\ket{g}$ and $\ket{e}$ states. The resonance position of $\ket{g}-\ket{e}$ transition is characterized by performing absorption imaging using this narrow line, the details of which can be found elsewhere \cite{Hu2021}. For driving the Rabi oscillation, we start from atoms at the ground state in the ODT and apply an on-resonance pulse with the power used in the Rydberg excitation, after which the ground state atom number is measured via absorption imaging on the $\ket{g}-\ket{i}$ transition within \SI{1}{\mu s}. This imaging time is short compared to the $\ket{e}$ lifetime ($\sim$~\SI{21}{\mu s}), thus minimizing the decay from $\ket{e}$ to $\ket{g}$ during the detection.

\begin{figure}[t!]
	\centering
	\includegraphics[width=0.5\textwidth]{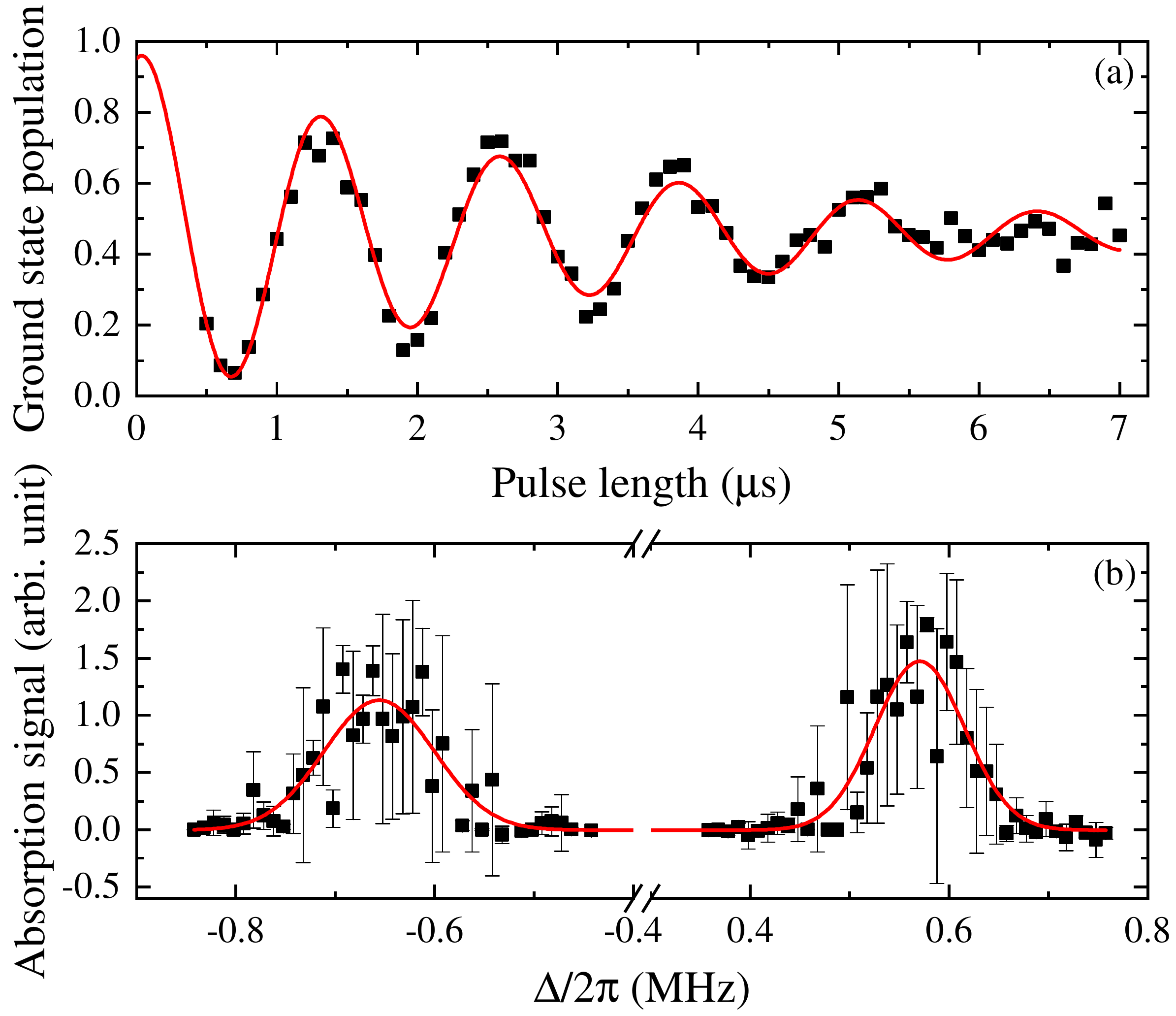}
	\caption{Calibrations of Rabi frequencies. (a) Rabi oscillation between $\ket{g}$ and $\ket{e}$. The solid lines are fits to a damped cosinusoidal function (see text) and the obtained fitting parameters are: $\Omega_{ge}/2\pi=$~\SI{784(5)}{kHz}, $\tau=$~\SI{3.0(3)}{\mu s}, and $t_0=$~\SI{0.05(2)}{\mu s}. (b) ATS caused by the UV light measured via absorption imaging on the red transition. The data is fitted to the analytic expression in Eq. \eqref{eq:ATS}, from which we obtain: $\Omega_{er}/2\pi=$~\SI{1235(13)}{kHz} and the UV detuning $\delta_{er}/2\pi=$~\SI{86(7)}{kHz}. }
	\label{fig:appendix}
\end{figure}

\begin{figure}[t]
	\centering
	\includegraphics[width=0.5\textwidth]{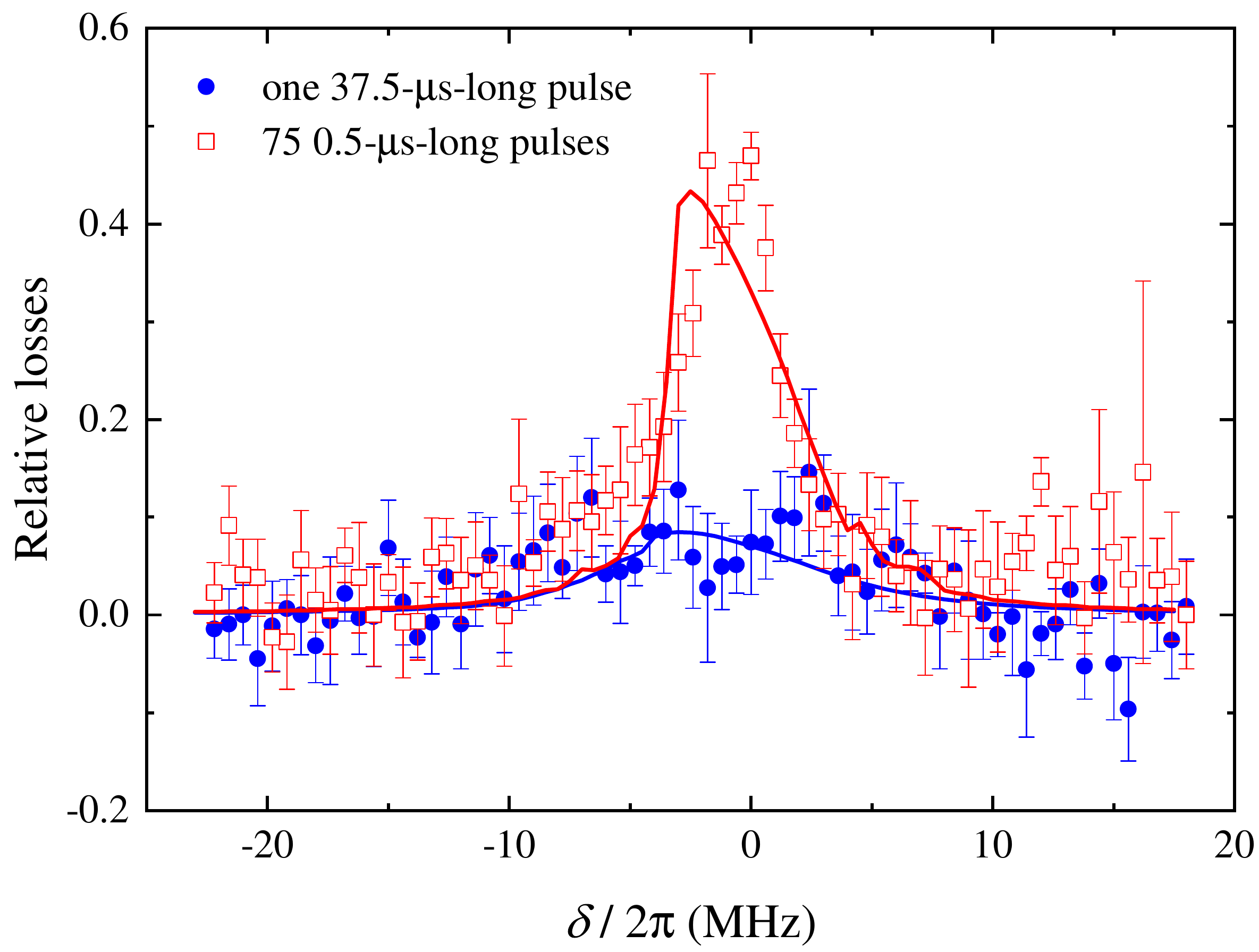}
	\caption{Comparison between the experimental data (blue and red points) from Fig. 1(b) of Ref. \cite{DeHond2018} and our mean-field calculation (solid curves). See text for descriptions.}
	\label{fig:Amsterdam}
\end{figure}

\begin{figure*}[t]
	\centering
	\includegraphics[width=\textwidth]{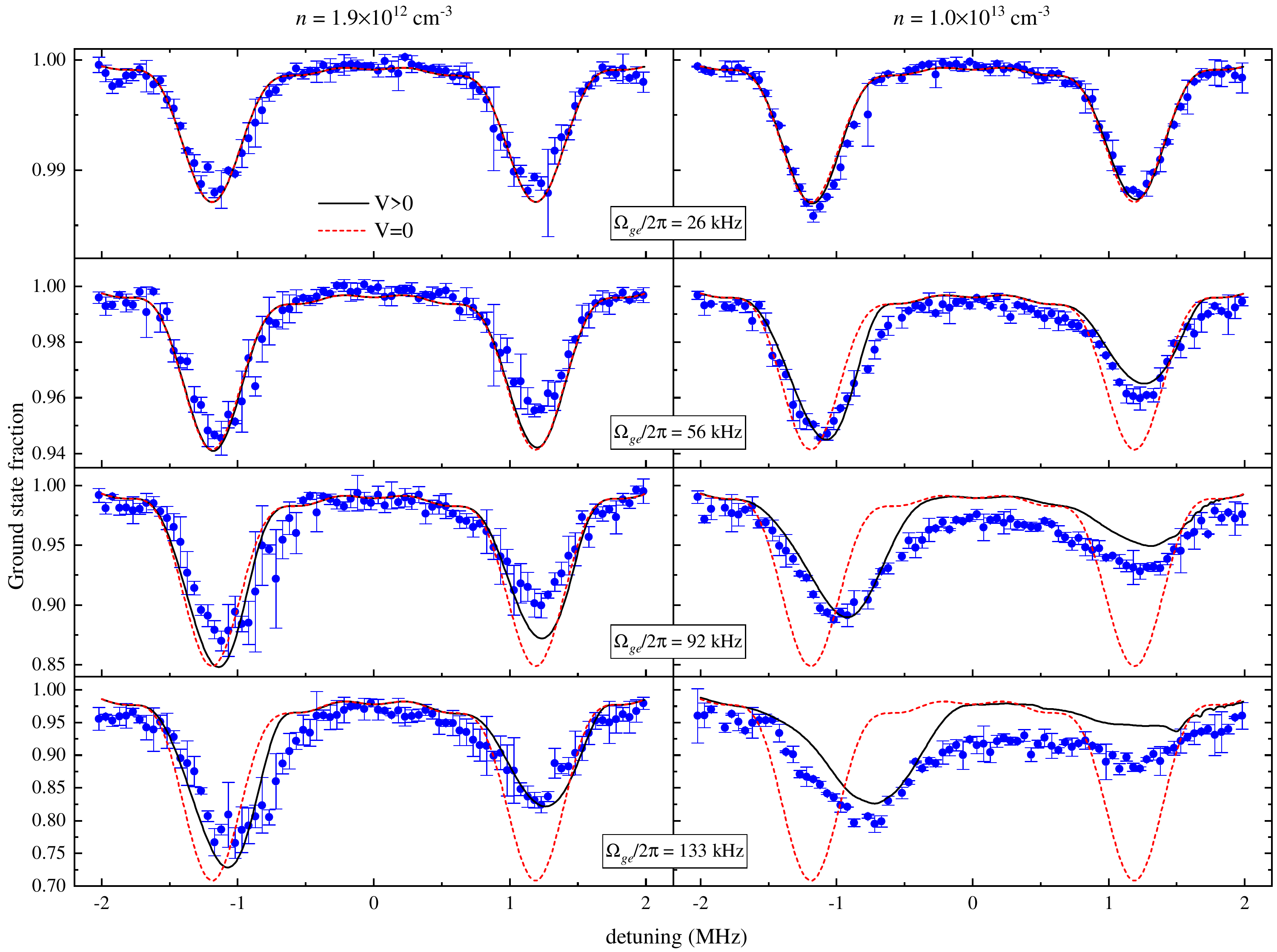}
	\caption{Comparison between the experimental data from Figs. 8 and 9 of Ref. \cite{DeSalvo2016} and our mean-field calculation. The black curves represent the case of including the Rydberg interaction term while the red dashed ones are free of interaction ($V=0$). See text for more details.}
	\label{fig:Rice}
\end{figure*} 

One example of the Rabi oscillation is shown in Fig. \hyperref[fig:appendix]{4(a)}. The data is fitted to a damped cosinusoidal function $A\exp[-(t-t_0)/\tau]\cos[\Omega(t-t_0)]+y_0$ to extract the oscillation frequency $\Omega$ and the decay time constant $\tau$. We introduce $t_0$ in the fits to account for the finite rising and falling edges of the square pulses used in the experiment ($\sim$\SI{50}{ns} for both). The detection fidelity is inferred to be about $96\%$.

The oscillation frequency $\Omega$ is interpreted as the Rabi frequency $\Omega_{ge}$ and the decay time constant is related to the dephasing term $\gamma_{ge}$ as $\gamma_{ge}=2/\tau-\gamma_e$ in the limit of $1/\tau\gg\gamma_e$. However, we observed a much smaller decay rate of the Rabi oscillations for smaller Rabi frequencies ($\sim$\SI{20}{\mu s}). Such a dependence of $\tau$ on the Rabi frequency suggests homogeneous dephasing effects and may result from the characteristic phase noise distribution of our PDH-locked laser \cite{Qiao2019a}, where only noises at frequencies around $\Omega_{ge}$ play the important role for the dephasing. The inhomogeneous Doppler shifts due to the finite temperature (with a Doppler width $\sim2\pi\times$~\SI{52}{kHz}) have significant effects only when the power-broadened linewidth is around the Doppler one.

\subsection{The UV-induced Aulter-Townes splitting}
\label{sec:ATS}
To characterize the Rabi frequency $\Omega_{er}$ and the two-photon detuning $\delta$, we measure the ATS of the $\ket{e}$ state via probing the absorption of a weak red beam by the atomic cloud in the presence of a strong UV light. The red probe light is overlapped with our blue imaging beam and we use a standard sequence for performing the red absorption imaging with a pulse of \SI{150}{\mu s} \cite{Hu2021}. The total absorption over the whole atomic cloud is measured as a function of the probe detuning, as shown in Fig. \hyperref[fig:appendix]{4(b)} for $n=23$. 

The spectrum is fitted to an analytic expression for the ATS [solid line in Fig. \hyperref[fig:appendix]{4(b)}] \cite{Fleischhauer2005}, which reads
\begin{equation}
	\label{eq:ATS}
	\sigma\propto\frac{\gamma_{ge}^2\gamma_{er}+\gamma_{er}\delta^2+\gamma_{ge}\Omega_{er}^2/4}{(\gamma_{er}\delta+\gamma_{ge}\Delta)^2+(\gamma_{er}\gamma_{ge}-\delta\Delta+\Omega_{er}^2/4)^2} \, .
\end{equation}
Here $\sigma$ represents the absorption cross section. The UV Rabi frequency $\Omega_{er}$ and the two-photon detuning $\delta$ is obtained from this fitting. The two dephasing terms $\gamma_{ge}$ and $\gamma_{er}$ are also free fitting parameters. In Fig. \hyperref[fig:appendix]{4(b)}, $\gamma_{ge}=2\pi\times0.08(2)$ is smaller than the dephasing rate obtained in Fig. \hyperref[fig:appendix]{4(a)} due to smaller $\Omega_{ge}$ used here. We obtain $\gamma_{er}=2\pi\times0.04(2)$, which agrees with the fitted dephasing in Table \ref{tab:interaction} for the low-density data. 

\section{Benchmarking the mean-field model} \label{sec:beachmark}

The mean-field model presented in Sec. \ref{sec:model} is benchmarked here with previously published experimental data from two different experiments, e.g. Refs. \cite{DeHond2018} and \cite{DeSalvo2016}. 

The authors in in Ref. \cite{DeHond2018} used the same treatment of the Rydberg-Rydberg interaction term as presented here and they simplified the four-level system to a three-level problem by adiabatically elimilating the intermediate level $\ket{e}$ due to a large one-photon detuning ($\sim2\pi\times$\SI{100}{MHz}). The experimental data in Fig. 1(b) of Ref. \cite{DeHond2018} is reproduced in Fig. \ref{fig:Amsterdam}, where the atomic losses were measured as a function of the two-photon detuning $\delta$ for two different excitation cases, e.g. a single long excitation pulse of \SI{37.5}{\mu s} (red squares) and 75 equally-separated short pulses of \SI{0.5}{\mu s} (blue circles). The latter pulse scheme is similar as what we implement here [see Fig. \hyperref[fig:appendix]{1(c)}]. Two theoretical curves modelling these two excitation schemes with parameters taken from Ref. \cite{DeHond2018} are also shown in Fig. \ref{fig:Amsterdam}, which agree quite well with the experimental data. 

The comparison between our mean-field calculations with the experimental data shown in Figs. 8 and 9 of Ref. \cite{DeSalvo2016} is presented in Fig. \ref{fig:Rice}, where atom losses in an optically trapped strontium gas were recorded in the same multi-pulse excitation scheme as we use here while scanning the red laser detuning with the UV frequency fixed. This atom-loss ATS spectrum was measured for four different $\Omega_{ge}$ at two atomic densities. A similar mean-field analysis was performed in Ref. \cite{DeSalvo2016} and described the data excellently, where the main difference from ours is that a different interaction term $V'=\frac{4\pi}{3}\sqrt{2C_6\Omega_{er}}n_0\rho_{rr}$ compared to Eqs. \eqref{eq:hamiltonian} and \eqref{eq: Interaction} was used. In despite of that, our mean-field calculation with the Hamiltonian \eqref{eq:hamiltonian} also agrees quantitatively well with the experimental data. Both interaction terms incorporate the short-range correlations induced by the Rydberg blockade by introducing a hardcore cutoff $R_b$, which was given by the blockade radius $(C_6/2\Omega_{er})^{1/6}$ in Ref. \cite{DeSalvo2016} and determined self-consistently in the model used here \cite{Helmrich2018}. 

\bibliographystyle{unsrt}
\bibliography{mylibrary}


\end{document}